\documentclass[aps,prl,twocolumn,superscriptaddress,showpacs,preprintnumbers]{revtex4}
\usepackage{graphicx}
\usepackage{amssymb}

\newcommand{\afc}       {antiferromagnetic}

\newcommand{\co}        {$^{59}$Co}
\newcommand{\ind}        {$^{115}$In}

\newcommand{\cecoin}    {CeCoIn$_5$}
\newcommand{\cerhin}    {CeRhIn$_5$}

\newcommand{\tc}     {$T_{\rm c}$}

\newcommand{\hctwo}     {$H_{\rm c2}(0)$}

\begin{document}


\preprint{LA-UR-06-2718}

\title{Microscopic evidence for field-induced magnetism in CeCoIn$_5$}
\author{B.-L. Young}
\altaffiliation[Current address:]{National Chiao Tung University,
Taiwan}
\author{R. R. Urbano}
\author{N. J. Curro}
\author{J. D. Thompson}
\author{J. L. Sarrao}
\affiliation{Condensed Matter and Thermal Physics, Los Alamos
National Laboratory, Los Alamos, NM 87545, USA}
\author{A. B. Vorontsov}
\affiliation{Department of Physics and Astronomy, Louisiana State
University, Baton Rouge, Louisiana 70803, USA}
\author{M. J. Graf}
\affiliation{Theoretical Division, Los Alamos National Laboratory,
Los Alamos, NM 87545, USA}

\date{\today}

\begin{abstract}
We present NMR data in the normal and superconducting states of
\cecoin\ for fields close to \hctwo$=11.8$ T in the $ab$ plane.
Recent experiments identified a first-order transition from the
normal to superconducting state for $H> 10.5$ T, and a new
thermodynamic phase below 290 mK within the superconducting state.
We find that the Knight shifts of the In(1), In(2) and the Co are
discontinuous across the first-order transition and the magnetic
linewidths increase dramatically. The broadening differs for the
three sites, unlike the expectation for an Abrikosov vortex lattice,
and suggests the presence of static spin moments in the vortex
cores. In the low-temperature and high-field phase the broad NMR
lineshapes suggest ordered local moments, rather than a long
wavelength quasiparticle spin density modulation expected for an
FFLO phase.

\end{abstract}


\maketitle

One of the most intriguing properties observed in Kondo lattice
systems is the emergence of unconventional superconductivity near a
quantum critical point (QCP).   By varying some external parameter
such as field or pressure, an \afc\ ground state can be tuned such
that the transition temperature goes to zero at the QCP.  As the
tuning parameter increases past the QCP, conventional Fermi-liquid
behavior is recovered below a characteristic temperature $T_{\rm
FL}$ \cite{YRSnature}. Superconductivity often emerges as the ground
state of the system for sufficiently low temperatures in the
vicinity of the QCP \cite{lonzarichCeIn3}. The heavy-fermion
superconductor \cecoin\ exhibits many properties typical of a Kondo
lattice system at a QCP.  In particular, $T_{\rm FL}$ appears to
vanish at the superconducting critical field $H_{\rm c2}(T=0)$ for
fields along the c axis, suggesting the presence of a field-tuned
QCP \cite{romanQCPCeCoIn5,taillefair}. This interpretation has
remained contentious because the ordered state associated with the
QCP is superconductivity rather than antiferromagnetism.  One
explanation is that an antiferromagnetic (AFM) phase is hidden
within the superconducting phase diagram, which is the genitor of
both the QCP and non-Fermi liquid behavior in the vicinity of
\hctwo. However, when the superconductivity is suppressed with Sn
doping, the QCP tracks \hctwo, and no magnetic state emerges in the
phase diagram, whereas pressure separates the QCP
\cite{ronningbauer}.

In fact, there is a field-induced state, which we will refer to as
the B phase, in the $H-T$ phase diagram of \cecoin\ that exists just
below \hctwo. The order parameter of the B phase could be either (1)
a different symmetry of the superconducting order parameter, (2) a
field-induced magnetic phase, or (3) a
Fulde-Ferrell-Larkin-Ovchinnikov (FFLO) superconducting phase
\cite{romanFFLO,andrea,bianchi2,radovan}.  The normal to
superconducting transition in this system has a critical point at
$(H,T) \sim (10.5 {\rm T}, 0.75 {\rm K})$, separating a second to
first order transition, and the B phase exists below a temperature
$T_0(H) \sim 290$ mK and is bounded by $T_c(H)$. NMR experiments
suggest the presence of excess quasiparticles associated with nodes
in the superconducting FFLO wavefunction
\cite{ff,lo,maki,kumagaiCeCoIn5}. However, recent NMR work by
Mitrovi\'{c} et al. disagrees with the original study, casting doubt
on the interpretation of this ordered phase as an FFLO state
\cite{mitrovic}. In this Letter we report detailed NMR spectra of
all three sites: the $^{115}$In(1), $^{115}$In(2) and $^{59}$Co, in
the normal and superconducting phases. Our data agree with those of
\cite{mitrovic}, and by comparing our spectra at the three sites, we
conclude that long-range order of local moments exists below $T_0$.
Therefore, the B phase is neither a different symmetry of the
superconducting order parameter, nor simply the FFLO state, but
rather a more complex field-induced magnetic state that may be
responsible for the QCP point at \hctwo.

We also find evidence for field-induced magnetism in the mixed state
(A phase) between \tc\ and $T_0$.  In this temperature and field
range, we find that the NMR Knight shift is discontinuous across the
first-order transition ($T_c(H)<750$ mK), and the spectra undergo a
dramatic magnetic broadening nearly one order of magnitude larger
than expected for orbital currents in a vortex lattice. The
broadening is different for the Co and In(1) sites, suggesting that
the origin of the magnetic broadening is a distribution of hyperfine
rather than orbital fields.  A likely source of hyperfine fields are
quasi-static spin moments within the vortex cores.

\begin{figure}
\includegraphics[width=\linewidth]{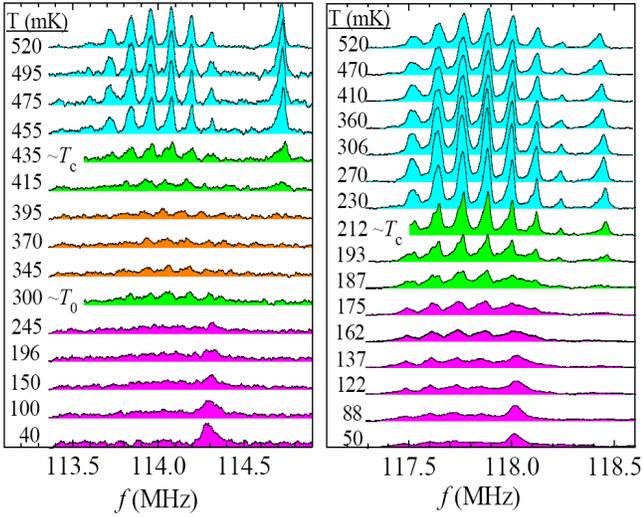}
\caption{\label{fig:spec} NMR spectra of \cecoin\ at 11.1T (left)
and at 11.485T (right). The series of  transitions at lower
frequency are the seven transitions of the \co, and the resonance at
higher frequency is the
$\left(-\frac{3}{2}\leftrightarrow-\frac{5}{2}\right)$ transition of
the \ind(1). The light blue shaded spectra are in the normal state,
the green spectra are within 20 mK of \tc\ ($T_0$), the orange
spectra are in the A phase, and the purple spectra are in the B
phase.}
\end{figure}

All of the NMR measurements were made on a single crystal of
\cecoin\ mounted with $H || a$. The orientation was verified to
within $\sim 1^\circ$ by observing the resonance frequencies of the
quadrupolar satellites of the In(1) ($^{115}I = 9/2$).  The sample
was mounted in the $^3$He-$^4$He mixture of a dilution refrigerator,
and the tank circuit was tuned by two fixed capacitors located close
to the coil. Spectra were obtained by summing several individual
spectra taken with low power at constant frequency intervals
\cite{GilClarkFFTSum}.  The temperature was monitored by a ruthenium
oxide resistor mounted close to the sample.  Heating of the sample
was minimized by reducing the pulse power to within less than 200 mW
for less than 20 $\mu$s. The field of the magnet was not
independently calibrated, so the Knight shift measurements were
shifted so that the normal state values extrapolated to those
measured previously \cite{CurroAnomalousShift}.

\begin{figure}
\includegraphics[width=\linewidth]{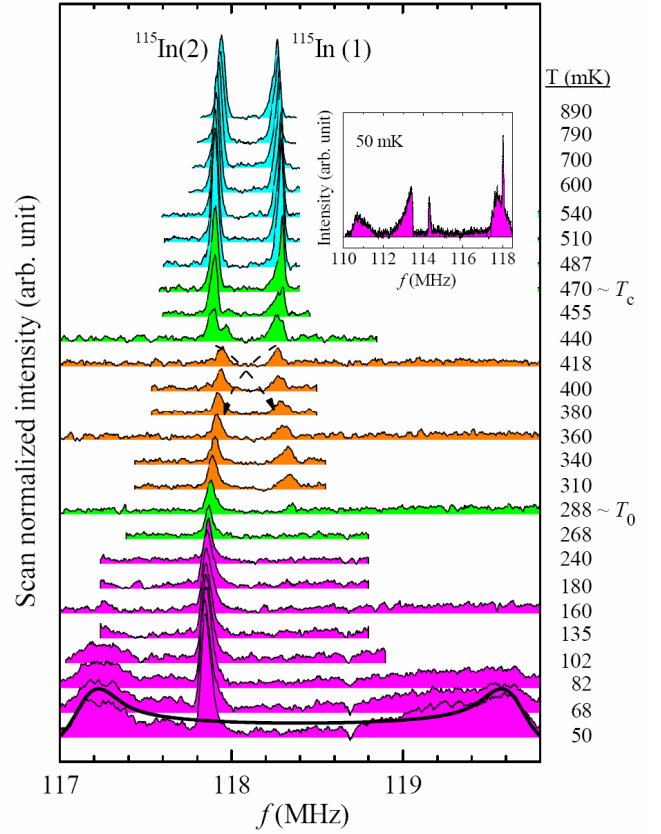}
\caption{\label{fig:spectra111} NMR spectra of In(1)
$\left(-\frac{5}{2}\leftrightarrow-\frac{7}{2}\right)$ and the
In(2)$_{||}$ $\left(-\frac{1}{2}\leftrightarrow-\frac{3}{2}\right)$
transitions in \cecoin\ at 11.1T. Note that the In(1) transition at
118.3 MHz in the normal state shifts down in frequency
discontinuously at \tc$\sim470$ mK, whereas the In(2)$_{||}$ shifts
up in frequency, as observed previously in lower fields
\cite{CurroAnomalousShift}. The broad double-peak structure between
117 and 119.5 MHz at 50 mK is the In(2)$_{||}$ spectrum, and the
solid line is a simulation as discussed in the text. INSET: the
spectrum at 11.485 T, showing the broadened In(2)$_{||}$ central
transition between 110 and 113 MHz.}
\end{figure}

\begin{figure}
\includegraphics[width=\linewidth]{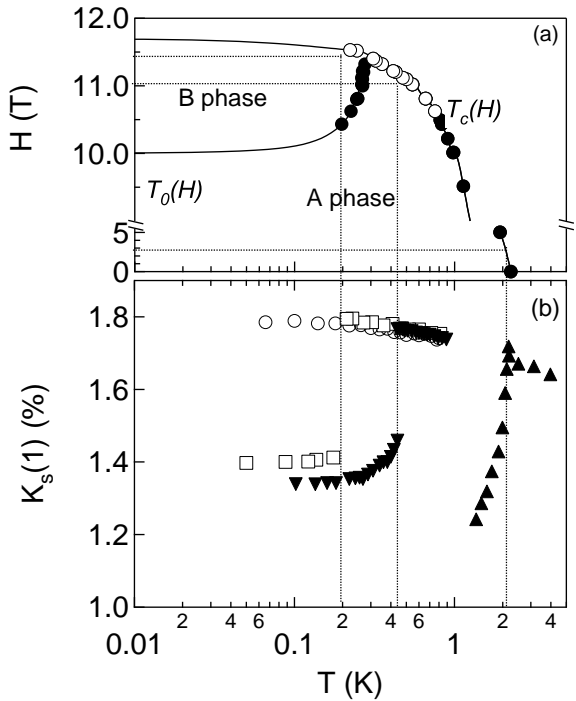}
\caption{\label{fig:shift} (a) The $H-T$ phase diagram, showing
first-order ($\circ$) and second-order ($\bullet$) transitions, from
\cite{tayama}. The solid lines are guides to the eye. (b) The Knight
shift of the In(1) at 11.8T ($\circ$), 11.485T ($\square$), 11.1T
($\blacktriangledown$), and 3.33T ($\blacktriangle$,
\cite{CurroAnomalousShift}).}
\end{figure}

Figure \ref{fig:spec} shows spectra of the Co and In(1) at two
different fields as a function of temperature (see Fig.
\ref{fig:shift}a). The In(1)
$\left(-\frac{5}{2}\leftrightarrow-\frac{3}{2}\right)$ transitions
at $\sim$114.7 MHz and $\sim$118.5 MHz shift to lower frequency
discontinuously  at $T_c$. We have confirmed that the resonances at
$\sim 114.3$ and $\sim 118.1$ MHz for $T<200$ mK are indeed the
In(1) by measuring several satellite transitions that show similar
shifts in the superconducting state. The quadrupolar splitting
between the Co and In(1) satellites remains temperature independent,
indicating that the discontinuity in the resonance frequency has a
magnetic origin, rather than a change in the charge configuration.
The absolute intensity of the spectra drops at \tc, an indication
that the sample is superconducting as the rf penetration is reduced.
Fig.~\ref{fig:spectra111} shows spectra of In(1) and In(2)$_{||}$
($\mathbf{H}$ parallel to the face of unit cell
\cite{CurroAnomalousShift}) at 11.1T.

Figure \ref{fig:shift}b shows the temperature dependence of the
In(1) Knight shift, $K_s$, as a function of temperature and field.
$K_s$ is determined from the first moment of the resonance, and we
have subtracted the temperature independent orbital shift $K_{\rm o}
= 0.13\%$ to obtain the spin contribution
\cite{CurroAnomalousShift}. We find a discontinuous jump in $K_s(T)$
at \tc, in agreement with bulk measurements at these fields, which
reflects the discontinuity in the superconducting gap at the first
order transition \cite{tayama,andrea}.

\begin{figure}
\includegraphics[width=\linewidth]{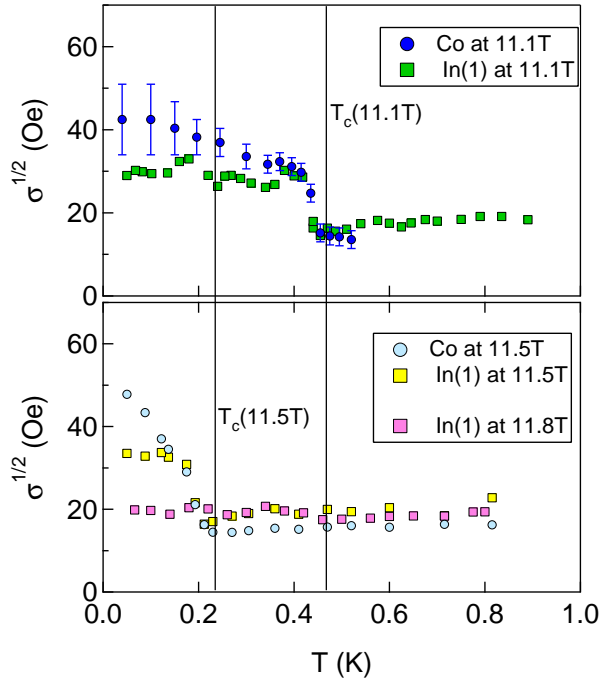}
\caption{\label{fig:linewidth} The second moments of the Co and
In(1) resonances at 11.1 T (upper) and at 11.485 T (lower).  The
normal state data at 11.8 T is included in the lower panel.}
\end{figure}

The spectra shown in Figs. \ref{fig:spec} and \ref{fig:spectra111}
clearly show a dramatic increase in the magnetic linewidths below
\tc. The linewidths of the Co and In(1) are shown as functions of
field and temperature in Fig. \ref{fig:linewidth}. The resonance
frequency in the superconducting state can be written as the sum of
three contributions: $\label{eqn:nmrfreq} f(\mathbf{r}) =
{}^{\eta}\gamma|\mathbf{H} + 4\pi \mathbf{M}_o(\mathbf{r})+
A\mathbf{M}_s(\mathbf{r})|$, where $^{\eta}\gamma$ is the
gyromagnetic ratio of the $\eta$ nucleus, $A$ is the hyperfine
coupling, $\mathbf{M}_{o(s)}(\mathbf{r})$ is the orbital (spin)
magnetization and $K_s= AM_s/H$. There are two sources of magnetic
broadening: a spatial distribution of $\mathbf{M}_o(\mathbf{r})$ or
$\mathbf{M}_s(\mathbf{r})$. In type II superconductors,  both are
spatially distributed due to the vortex lattice, and hence the NMR
spectrum develops a characteristic lineshape in the mixed state (A
phase), which is typically dominated by $\mathbf{M}_o(\mathbf{r})$
\cite{MacLaughlin,CurroSlichter}. However, the broadening we observe
occurs in the A phase and changes little in the B phase. This result
is surprising, since \textit{a priori} one would expect an extra
broadening due to $\mathbf{M}_s(\mathbf{r})$ in the FFLO phase
\cite{grafknight,graf}. In fact, the vortex contribution,
$\mathbf{M}_o(\mathbf{r})$, should be negligible at these fields.
The second moment of the Abrikosov vortex lattice field distribution
with a Ginzburg-Landau parameter $\kappa=\lambda/\xi \approx 60$ and
orbital limiting field $H_{c2}^{\rm o}\approx35$ T is
$\sqrt{\sigma_{\rm orb}}\approx 12$ Oe at these fields
\cite{cecoin5pendepth,tayama,brandt}. Convolving this result with
the intrinsic normal state linewidths ($\sqrt{\sigma_{\rm n}}\sim20$
Oe), gives a net change of $\sim3$ Oe. Clearly, as seen in Fig.
\ref{fig:linewidth}, the magnetic broadening observed is much too
large to explain with a conventional Abrikosov vortex lattice.
Furthermore, the broadening at the Co site is nearly twice that at
the In(1) site. If the broadening mechanism were from orbital
supercurrents or spin-polarized quasiparticles in the domain walls
of the FFLO state, then the response at the Co and In(1) would be
identical.  The only way to understand our results is a distribution
of $\mathbf{M}_s(\mathbf{r})$ due to local moments, which gives a
different response for different hyperfine couplings unique to each
nuclear site.

We propose that this distribution of spin polarization arises from
magnetic order in the vortex cores, as has been found in the high
temperature superconductors \cite{lakescience,kumagaiAFVortex}.
Since the superconducting order parameter vanishes in the cores, it
is plausible that competing orders may be stabilized \cite{demler}.
\cecoin\ becomes AFM with only a few percent Cd doping, which
indicates that this material lies close to an AFM instability
\cite{FiskCddoping}. Indeed, recent neutron measurements found an
enhancement of the vortex lattice form factor consistent with spins
in the cores \cite{debeer}, and magnetization measurements as a
function of field reveal a strong paramagenetic contribution even in
the mixed state of this material \cite{tayama}. Such an effect has
been observed in other heavy-fermion materials, and may be
associated with a paramagnetic response of local f moments in the
vortex cores \cite{mixedparamagneticchipapers}.

Below $T_0$, the response of all three sites differs dramatically.
Figures \ref{fig:spec} and \ref{fig:spectra111} show spectra of the
In(1) and In(2)$_{||}$ at 11.1 T.  The In(1) and Co spectra change
little across the $T_0 \sim 290$ mK transition at 11.1T, whereas the
In(2)$_{||}$ signal disappears below $T_0$ and then reappears below
$\sim$100 mK with a broad double-peak structure over a range of
$\sim$2.5 MHz. We have confirmed that this spectrum is the
In(2)$_{||}$ by checking the response of a different satellite
transition (see Fig. \ref{fig:spectra111} INSET). Similar features
were observed in \cerhin\ in the AFM state, where the In(1) lines
remained sharp while the In(2) spectra developed a broad powder
pattern-like spectrum as a result of the incommensurate magnetic
structure \cite{CurroCeRhIn5}. Such an effect cannot be explained by
a long-wavelength modulation of $M_{s}(\mathbf{r})$ as expected in
an FFLO state, or a change of the order parameter symmetry. In
either case, the wavelength of the modulation should be on the order
of either the coherence length, $\xi$, or the Fermi wavevector
mismatch, $1/|\mathbf{k}_{{\rm F}\uparrow}-\mathbf{k}_{{\rm
F}\downarrow}|$. Both of these length scales exceed the unit cell
length, implying that the response of the Co, In(1) and In(2) should
be similar. If there were static order of Ce moments, then because
of their particular site symmetries the Co and In(1) can remain
relatively sharp whereas the In(2) can experience large hyperfine
fields \cite{currohyperfine}. A possible magnetic structure that
satisfies these requirements is $\mathbf{Q}=(\pi/a-\delta, \pi/a,
\pi/c)$, where $2\pi/\delta$ is the wavelength of the
incommensuration and the Ce spins
$\mathbf{S}_0~||~\mathbf{H}~||~\hat{a}$. In this case the isotropic
components of the hyperfine field at the In(1) and Co sites
vanishes, but at the In(2) the hyperfine field has components either
parallel or antiparallel to $\mathbf{H}$. The solid line in Fig.
\ref{fig:spectra111} shows the expected lineshape for a sinusoidal
variation of $H_{\rm hyp}$ with magnitude 1.3 kOe, which has been
convolved with a Gaussian with width 100 kHz. We do not have
independent information to determine either $\delta$ or $S_0$, since
$H_{\rm hyp}\propto S_0\delta$. The onset of long-range magnetic
order also explains why the In(2)$_{||}$ signal disappears just
below $T_0$, since the combination of critical slowing down and the
large hyperfine fields leads to a fast spin-echo decay time, $T_2$,
wiping out the NMR signal \cite{CurroNJ:Inhlfs}. When the magnetic
order becomes static, $T_2$ becomes longer and the signal recovers,
but the large static hyperfine field shifts the resonance frequency.

A possible explanation for understanding these results is that the
field-induced magnetism in the vortex cores becomes correlated
between the vortices below $T_0$.  The isostructural compound
\cerhin\ exhibits field induced magnetism under pressure
\cite{tuson}. Comparison of the pressure dependent phase diagrams of
these two materials suggests that \cecoin\ is nearly identical to
\cerhin\ under a pressure of 1.6-2.3 GPa, exactly in the vicinity of
the pressure where \cerhin\ exhibits field-induced magnetism
\cite{sidorov}. Furthermore, the $H-T$ phase diagram of \cerhin\ is
nearly identical to that of \cecoin, except that in \cerhin\ the
field-induced magnetism persists above \hctwo, whereas in \cecoin\
there is no sign of magnetism in the normal state. We cannot rule
out the existence of an FFLO state, or whether the long-range
magnetism coexists with the FFLO order. Nevertheless, local moment
magnetism clearly competes with Kondo screening and with
superconductivity, so magnetism may emerge naturally where the
superconductivity is suppressed within the vortex cores or the nodal
planes of the FFLO phase. This interpretation offers a consistent
explanation of the non-Fermi liquid behavior associated with the QCP
at \hctwo, where the observed field-induced magnetism apparently
exists only within the superconducting phase.

We thank A. Balatsky, A. Bianchi, L. Boulaevskii, M. Nicklas, R.
Movshovich, T. Park, and F. Ronning for enlightening discussions. A.
B. V. was supported by the Louisiana Board of Regents. This work was
performed at Los Alamos National Laboratory under the auspices of
the U.S. Department of Energy.


\end{document}